%Paper: astro-ph/9508110
%From: Yoshiaki Sofue <sofue@sof.mtk.ioa.s.u-tokyo.ac.jp>
%Date: Thu, 24 Aug 95 18:18:20 JST
%Date (revised): Wed, 25 Oct 95 14:00:57 JST

\magnification=1200
%%%%%%%%%%%%%%%
%% Definitions
%%%%%%%%%%%%%%%
\def\lv{$(l, V)$}
\def\bv{$(b, V)$}
\def\lb{$(l, b)$}
\def\v{\vskip 2mm}
\def\vv{\v\v}
\def\vvv{\v\v\v}
\def\noi{\noindent}
\def\no{\noindent}
\def\title#1{\centerline {\bf #1}\vskip 8mm}
\def\author#1{\centerline{#1}\vskip 2mm}

\def\abstract#1{{\centerline{\bf Abstract}}\vskip 2mm {#1}}

\def\section#1{\vskip 8mm \noi {\bf #1} \vskip 8mm}
\def\subsection#1{\vskip 6mm \noi {\bf #1} \vskip 6mm}

\def\sub#1{\vskip 6mm \noi {\bf #1} \vskip 4mm}

\def\r{\hangindent=1pc  \noindent}

\def\ce{\centerline}
\def\endpage{\vfil\break}
\def\kms{km s$^{-1}$}

\def\deg{$^\circ$}

\def\vlsr{$V_{\rm LSR}$}

\def\Vlsr{V_{\rm LSR}}

\def\Vrot{V_{\rm rot}}

\def\tb{$T_{\rm B}$}

\def\kms{km s$^{-1}$}
\def\Msun{M_{\odot \hskip-5.2pt \bullet}}

\def\Deg{^\circ}
\def\deg{$^\circ$}

\def\htwo{H$_2$}

\def\ta{$T^*_{\rm A}$}
\def\tb{$T_{\rm B}$}

%%%TEXT%%%

\ce{\bf Galactic Center Molecular Arms, Ring and Expanding Shells. I}
\ce{\bf Kinematical Structures in Longitude-Velocity Diagrams}

\vvv
\vvv
\ce{Yoshiaki SOFUE}

\vv
\ce{Institute of Astronomy, The University of Tokyo}
\ce{ Mitaka, Tokyo 181, Japan}
\ce{ sofue@mtk.ioa.s.u-tokyo.ac.jp}

\vv
\ce{({\it To appear in PASJ Vol. 47, No.5})}
\vvv
\vvv
\ce{\bf Abstract}  \vv

Analyzing the $(l, b, \Vlsr)$ data cube of $^{13}{\rm CO}(J=1-0)$
line emission obtained by Bally et al,
we have investigated the molecular gas distribution and
kinematics in the central $\pm 1\Deg ~(\pm 150$ pc) region of the Galaxy.
We have applied the pressing method to remove the local- and foreground-gas
components at low velocities in order to estimate the intensity more
quantitatively.
Two major dense molecular arms
have been identified in longitude-radial velocity \lv\ diagrams as
apparently rigidly-rotating ridges.
The ridges are spatially identified as two  arms,
which we call the Galactic Center molecular Arms (GCA).
The arms compose a rotating ring of radius 120 pc (the 120-pc
Molecular Ring), whose inclination is  $i \simeq 85\Deg$.
The Sgr B molecular complex is associated with GCA  I, and
Sgr C complex is located on GCA II.
These arms are as thin as 13 to 15 pc, except for vertically extended
massive complexes  around Sgr B and C.
The \lv\ behavior of the arms  can be qualitatively
reproduced by a model which assumes spiral arms of gas.
Assuming a small pitch angle for the arms, we tried to deconvolve
the \lv\ diagram to a projection
on the galactic plane, and present a possible face-on CO map
as seen from the galactic pole, which also reveals a molecular ring and arms.
We have estimated  masses of these molecular features, using the most
recent value of  the CO-to-\htwo\ conversion factor taking into
account its metallicity dependency and radial gradient in the Galaxy.
The estimated molecular masses and kinetic energy  are about
a factor of three smaller than those reported in the literature
using the conventional conversion factor.

\vv\no{\bf Key words}: Galactic center -- Galaxy -- Interstellar matter
--  Molecular gas -- Spiral arms.

\vvv
\section{1. Introduction}

The galactic center region has been extensively observed in the molecular
lines, particularly in the CO line emission
(Oort 1977; Scoville et al 1974; Liszt 1988; Liszt and Burton 1978, 1980;
 Burton and Liszt 1983, 1993; Brown and Liszt 1984; Heiligman 1987;
Bally et al 1987; 1988; Genzel and Townes 1987; Stark et al 1989;
G{\"u}sten 1989).
Besides the 4-kpc molecular ring, the CO emission is strongly
concentrated in the central a few degree (Dame et al 1987).
Moreover, the molecular gas in the central region has a strong
concentration within $|l|<1\Deg$ (150 pc) where the majority of the nuclear
disk gas is confined (Scoville et al 1974; Bally et al 1987; Heiligman 1987).
This high  concentration of dense interstellar matter in a small region
is also clearly visible in the far IR emission (e.g., Cox and Laureijs 1989)
and in the CII emission (Okuda et al 1989).
The radio continuum emission also indicates a highly concentrated
nuclear disk of ionized gas (Altenhoff et al 1978; Handa et al 1987).
On the other hand, the region between galactocentric distances $\sim 200$ pc
($l\sim 1\Deg4$) and $\sim 2$ kpc (15\deg) appears almost empty in the CO
emission  (Bally et al 1987; Knapp et all 1985).

The total molecular mass in the $|l|<1\Deg$ region estimated from the
CO emission  amounts to $ \sim 1.4  \times 10^8 \Msun$ for a traditional
CO-to-\htwo\ conversion factor, or, more probably,
$\sim  4.6 \times 10^7 \Msun$ for a new conversion factor (see section 3).
On the other hand, the HI mass within the 1.2 kpc tilted disk ($l<8\Deg$)
is only of several $10^6 \Msun$ (Liszt and Burton 1980).
Hence, we may consider that the central $\sim 1$ kpc region is dominated
by the molecular disk of $\sim 150$ pc ($\sim 1\Deg$) radius, outside
of which the gas density becomes an order of magnitude smaller.

Various molecular gas features in the central $\sim100-200$  pc region
have been discussed by various authors, such as a disk
related to the 1.2 kpc tilted rotating disk
(Liszt and Burton 1980; Burton and Liszt 1992),
molecular rings and spiral arms of a few hundred pc scale
(Scoville et al 1974; Heiligman 1987; Bally et al 1987),
and the expanding molecular ring of 200 pc radius
(Scoville 1972; Kaifu et al 1972, 1974).
On the other hand, Binney et al (1991) have modeled the ``expanding-ring
feature" or the ``parallelogram'' on the \lv\ (longitude-velocity) plot
in terms of  non-circular kinematics of gas by
a closed orbit model  in a bar potential.
It is known that the gas in this parallelogram shares
only a small fraction of the total molecular mass in the galactic center:
the majority of the gas composes more rigid-body like features in the
\lv\ plots.

The CO gas in the central 100 - 200 pc regions in nearby galaxies have been
observed by high-resolution mm-wave interferometry,
and their distribution and kinematics have been extensively
studied (e.g., Lo et al 1984; Ishiguro et al 1989; Ishizuki et al 1990a,b).
The central gas disks of galaxies appear to comprise
spiral arms or circum-nuclear rings of a few hundred pc size.
Such a gaseous behavior can be reproduced to some extent by theoretical
simulations of accretion of gas clouds in a central
gravitational potential (e.g.,  Noguchi 1988; Wada and Habe 1992).

In this paper, we revisit the major part of the nuclear molecular disk
($|l|<\sim 1\Deg$) by analyzing the  molecular line data in the premise
that the nuclear disk may comprise accretion ring or spiral structures
similar to those found in external galaxies.
In this paper we reanalyze the data cube of the
$^{13}{\rm CO}~(J=1-0)$-line emission observed by Bally et al (1987)
with the 7-m off-set Cassegrain telescope of the Bell Telephone Laboratory.
The distance to the Galactic Center is  assumed to be 8.5 kpc throughout this
paper.

\section{2. Longitude-Velocity \lv\  Diagrams}

\sub{2.1. Data}

The angular resolution of the observations with the Bell-Telephone 7-m
antenna at $^{13}$CO line was 1$'.7$.
The data used here are in a $(l, b, \Vlsr)$ cube in FITS format.
The cube covers an area of $-1\Deg.1 \le l \le 0\Deg.92,~-21' \le b \le 17'$,
or 300 pc $\times$ 94 pc region for a 8.5 kpc distance.
The velocity coverage is $-250 \le \Vlsr \le 250$ \kms.
The cube comprises 127, 39, and 183  channels
at 1$'$,  1$'$, and 2.75 \kms\ intervals, respectively,
%%%CHECK below
We also  use the CS line data in a $(l,b, V)$
cube with dimensions 151, 42, and 163 channels at intervals
2$'$, 1$'$, and 2.75 \kms, which covers an area of
$-1\Deg \le l \le 4\Deg$, $-25' \le b \le 16'$, and $-250\le \Vlsr \le 190$
\kms.
The intensity scale of the data is the main-beam antenna temperature
approximately equivalent to brightness temperature in Kelvin.
The observational details are described in Bally et al (1987, 1988).
We made use of the AIPS and IRAF software packages for the reduction.

\sub{2.2. Subtraction of the Local and Foreground Components}

In order to analyze  molecular gas features in the Galactic Center
region, we first subtract contaminations by local and foreground
molecular clouds at low velocities.
Since $^{13}$CO line is optically thin for
the foreground clouds, the contaminations appear as emission
stripes superposed on the galactic center emission,
The subtraction of foreground emission is essential when we derive the mass
and kinetic energy of molecular gas features.
Such a ``cleaning"  also helps much the morphological recognition of features
on the \lv\ and  \bv\ diagrams.

Fig. 1a shows an \lv\ diagram averaged in a latitude range of
$-17' \le b \le 12'$.
The diagram is strongly affected by ``stripes'' at a low
velocities elongated in the direction of
longitude with narrow  velocity widths, including the 3-kpc expanding
arm at $-52$ \kms.
In order to eliminate these stripes, we applied the ``pressing method''
as developed for removing  scanning effects
in raster scan observations (Sofue and Reich 1979).
We briefly describe this method below.

\ce{--Fig. 1a, b --}

The original \lv\ map M$_0$ is trimmed by
$-70 \le \Vlsr \le 50$ \kms\ to yield M$_1$
where the local and foreground gas contribution is significant.
The trimmed map M$_1$ is  smoothed only in the $V$ direction by
5 channels (14 \kms) using a boxcar or Gaussian smoothing task, yielding M$_2$.
Smoothed map M$_2$  is subtracted from M$_1$  to yield M$_3 ~(=M_1 - M_2)$.
Map M$_3$ is then smoothed only in $l$ direction by 20 channels (20$'$)
(boxcar or Gaussian) to yield M$_4$.
This M$_4$ map approximates the contribution from the local gas that is
dominated by elongated features in the longitudinal direction.
We then subtract M$_4$ from M$_1$ to obtain M$_5~(=M_1-M_4)$.
This M$_5$ is, thus, a map in which the local gas contribution has been
roughly subtracted.
M$_5$ is then smoothed in $V$ direction by 5 channels.
Then we replace M$_2$ by this smoothed map, and repeat the above procedures
twice (or more times) until we obtain the second (or $n$-th) M$_5$.
Finally, the $-70 \le \Vlsr \le 50$ \kms\ part of the original map M$_0$ is
replaced by M$_5$ to yield M$_6$.
Now, we have a ``pressed'' map M$_6$ in which corrugations due to local
gas clouds have been removed out.
Fig. 1b shows the thus obtained map M$_6$ for the same \lv\ diagram as
in Fig. 1a.

We have applied this algorithm (the pressing method) to all \lv\ and \bv\
diagrams in the cube, and created a new $(l, b, V)$ cube, which is almost
free from local and foreground contaminations.
In the present paper we use this new cube.
We also applied the pressing method to remove scanning effects, which
had originated during the data acquisition,
in every diagram such as intensity maps in the $(l,b)$ space.

By comparing the original and the thus `pressed' maps,
we estimated the contribution of the local/foreground emission to be
5\% of the total emission, and 9\% of the emission with
$|\Vlsr|<100$ \kms.
Thus, without the subtraction, the mass and energetics  would
be overestimated by about  5 to 9\%.
Moreover, if the gas out of the disk component at $|b|>10'$ is concerned,
this local contribution would amount to more than 10\%.
Hence, the subtraction of the foreground emissions
is crucial in a quantitative discussion of the features discussed in this
paper.

\sub{2.3. ``Arms'' in Longitude-Velocity \lv\ Diagrams}

Fig. 2 shows \lv\ diagrams near the galactic plane
averaged in  $4'$ latitude interval
after subtraction of the local/foreground components.
Various features found in these diagrams have been discussed
in Bally et al (1987, 1988).
In this paper we highlight continuous features (ridges) traced
in the \lv\ diagrams.
The major structures of the ``disk component''
at low latitude ($|b|<\sim 10'$=25 pc)
are  ``rigid-rotation'' ridges, which we call  ``arms''.
Fig. 3 illustrate these ridges (arms) which can be identified in
the diagrams as coherent structures.
In  Table 1 we summarize the identified features, and describe below
the individual arms.
Heiligman (1987) has used these ridges to derive a rigid rotation curve.
At higher latitudes ($|b|>\sim 10'$) the so-called
expanding ring features at high velocities ($|\Vlsr >100$ \kms), which
will be discussed in a separate paper.

\ce{-- Fig. 2 --}
\ce{-- Fig. 3 --}
\ce{-- Table 1 --}

\sub{2.3.1. Arm I}

The most prominent \lv\ arm is found as a long and straight
ridge, slightly above the galactic plane at $b\sim 2'$,
which runs from \lv=($0\Deg.9, 80$  \kms) to ($-0\Deg.7, -150$ \kms),
and extends to  ($-1\Deg.0, -200$ \kms).
This arm intersects the line at $l=0\Deg$ at negative velocity
$\Vlsr=-40$ \kms, indicating that the gas is approaching us at $l=0\Deg$.
We call this ridge Arm I.
A part of this arm can be traced also below the galactic plane at $b=-0.1$\deg,
running from \lv=(0.8\deg, 60 \kms) to ($0\Deg.1, -20$ \kms).
Its positive longitude part is connected to the dense molecular
complex Sgr B, which is extended both in space and velocity, from
$b=-0.25$ to 0.07\deg\ and \vlsr=20 to 100 \kms.

\sub{2.3.2. Arm II}

Another prominent arm is seen at negative latitude at $b\sim -6'$,
running from \lv=(0\deg.1, 60 \kms) to $(-0\Deg.6, -80$ \kms).
We call this ridge Arm II.
It is bent at $l\sim 0\Deg.1$ and appears to continue to
\lv=(1\deg, 100 \kms), and merges with Arm I at the Sgr B complex region.
The negative longitude part also merges with Arm I, and is connected to the
Sgr C complex.
Arm II intersects  $l=0\Deg$ at positive velocity of $\Vlsr=50$ \kms.

\sub{2.3.3. Arms III and IV}

At positive latitude ($b\sim 0\Deg.01$ to $0\Deg.2$),
another arm can be traced running from
\lv=(0\deg, 140 \kms) to ($-0\Deg.15$, 10 \kms).
Its counterpart to the negative longitude side appears to be present
at \lv=($-0\Deg.45, -120$ \kms) to ($-0\Deg.55, -180$ \kms).
We call this ridge Arm III.
Bally et al (1988) called this the ``polar arc'', and discussed its
connection to Sgr A.

A branch can be traced from \lv=(0\deg.1, 60 \kms) to
($0\Deg, -20$ \kms), apparently being bifurcated from Arm II at $l\sim
0\Deg.1$.
This ridge intersects $l=0\Deg$ at negative velocity ($\Vlsr -50$ \kms).
We call this ridge Arm IV.

\sub{2.4. ``Rigid-rotation'' in \lv\ Plane and ``Arms and Ring'' in the
Galactic Plane}

We emphasize that ``rigid-rotation'' ridges  in \lv\ diagrams
for edge-on galaxies, whose rotation curves are usually flat,
are generally interpreted as due to spiral arms and rings.
Indeed,  the rigid-rotation ridge in the CO  \lv\ diagram
of the Milky Way is identified with the 4-kpc molecular ring
(e.g., Dame et al 1987; Combes 1992).
Many edge-on spiral galaxies like NGC 891  are found to show
similar \lv\ ridges in HI and CO, which are also interpreted to be
spiral arms and rings (e.g., Sofue and Nakai 1993, 1994).

The circular rotation velocity  as defined by
$V_{\rm rot}=(R \partial \Phi / \partial R)^{1/2}$, remains
greater than at least 150 \kms\ from the nuclear few pc region till the
1 kpc radius region (Genzel and Townes 1987).
Here,  $\Phi$ is the potential and $R$ is the distance from the nucleus.
Hence, the actual rotation  should not be rigid at all:
The rigid-rotation ridges in the \lv\ plane such as  Arms I to IV
in the Galactic Center can thus be more naturally attributed to real arms
and rings.

\section{3. Intensity Distribution and the Galactic Center Arms}

\sub{3.1. Intensity Maps and Masses}

Fig. 4a shows the total intensity map integrated over the full range of
the velocity ($-250 \le \Vlsr \le 250$ \kms).
This map is about the same as that presented by Stark et al (1989),
except that the local gas has been removed.
Fig. 4b shows the same in grey scale and a that with the  vertical
scale in $b$ direction enlarged twice.

\ce{-- Fig. 4 --}

First of all the intensity map can be used to obtain the molecular mass.
However, the conversion of the CO intensity
to \htwo\ mass  is not straightforward.
We  have recently studied the correlation  of
the conversion factor $X_{12}$ for the $^{12}$CO$(J=1-0)$ line
with the metal abundance in galaxies (Arimoto et al 1994).
We have obtained a clear dependency of $X$ on the galacto-centric
distance $R$ within individual galaxies, which is almost equivalent to the
metallicity dependence. For the Milky Way we have
$$ X_{12}(R)=0.92 (\pm 0.2)\times 10^{20}{\rm exp} (R/R_{\rm e})$$
where  $R_{\rm e}=7.1$ kpc is the scale radius of the disk.
Applying   this relation to the Galactic center, we obtain
a conversion factor at the Galactic center as
$X_{12}=0.92(\pm 0.2)\times10^{20}~{\rm [H_2~cm^{-2}/K~kms~s^{-1}]}$,
about one third of the solar vicinity value.
We then assume that the $^{12}$CO and $^{13}$CO intensities are
proportional, and  estimate the ratio by averaging observed intensity ratios
for the inner Galaxy at $l \le 20\Deg$ (Solomon et al 1979);
$I_{\rm 12CO}/I_{\rm 13CO} \simeq 6.2\pm 1.0$.
Then, we obtain a conversion factor for the $^{13}$CO line intensity
in the galactic center region as
$X_{13}(R=0)\simeq 5.7 \times 10^{20} ~{\rm [H_2~cm^{-2}/K~kms~s^{-1}]}$,
and we use this value throughout this paper.

The correction factor from the H mass to real gas mass is given
by  $\mu=1/X=1.61$, where $X$ is the hydrogen abundance in weight.
Here, the following relation has been adopted (Shaver et al 1983):
  $Y=0.28+(\Delta Y/\Delta Z) Z  = 0.34$ in weight, where
$Z=0.02$ is the abundance of the heavy elements and
 $\Delta Y/\Delta Z=3$ is the metallicity dependence
of the helium abundance $Y$ in the interstellar matter, and so,
the hydrogen abundance is $X=0.62$.
So, the surface mass density of molecular gas after correction
for the mean weight of gas is given by
$$ \sigma \sim 14.6 (\pm 3) ~ I/\eta~[ \Msun {\rm pc}^{-2}], \eqno(2) $$
where
$$ I\equiv\int T_{\rm A}^* dv {\rm~ [ K~ km~s^{-1}}] \eqno(3) $$
 is the integrated intensity of $^{13}{\rm CO}(J=1-0)$ line emission
and $\eta=0.89$ is the primary beam efficiency of the antenna.
The total mass of molecular gas (including He and metals) can be estimated by
$$ M~ [\Msun]= 14.6 \int I/\eta dx dy~ [{\rm K~km~s^{-1}~pc^2}]. \eqno(4) $$

%14.6==43.5x0.337(revised value on Sep 5, 1994)
% \eta=0.89 primary antenna efficiency (Bally et al 1987)

Using these relations, we have estimated the total molecular gas mass
in the observed area ($-1\Deg.0 \le l \le 0\Deg.92,~ -21' \le b \le 17')$
after removing the local and foreground contribution to be
$ 4.6  (\pm 0.8) \times 10^7 \Msun$.
We have also estimated the total molecular mass of the ``disk'' component,
which comprises most of the ridge-like features in the \lv\ diagrams,
excluding the expanding ring feature (or the parallelogram)
at high velocities ($|\Vlsr| >\sim 100 - 150$ \kms).
The disk component has the mass $3.9\times10^7 \Msun$, which is 85\% of the
 total in the observed region.
On the other hand, the expanding ring (or the parallelogram)
shares only $6.7\times 10^6 \Msun$ (15\%) in the region at $|l|<1\Deg$.

\sub{3.2. Ring and Arms in Intensity Maps}

In order to clarify if each of the arms traced in the \lv\ diagrams (Fig. 1-3),
particlurly Arms I and II, is a single physical structure in space, we have
obtained
velocity-integrated  intensity map in the \lb\ plane for each of the arms.
Thereby, we integrated the CO intensity in the velocity ranges as
shown in Fig. 5 individually for Arms I and II.
Fig. 6 show the integrated intensity maps corresponding to
Arms I and II, together with a summation of I and II.
In Table 1 we summarize the derived parameters.

\ce{-- Fig. 5 --}

\ce{-- Fig. 6 --}

\sub{3.2.1. Galactic Center Arms I, II}

In the intensity map, Arm I can be traced as a single, thin arc-like arm
from $l=0\Deg.9$ near the Sgr B complex toward negative longitude
at $l=-1\Deg.0$.
We call this spatial arm the Galactic Center Arm I (GCA I).
The angular extent is as long as $1\Deg.9$ (280 pc) in the longitudinal
direction, whereas the thickness in the $b$ direction is as thin as
 $\sim 5'$ (13 pc; see section 3.2.2).
The Sgr B molecular complex is much extended in the $b$ direction by
about $0\Deg.4$ (60 pc),  and composes  a massive part of the arm.
A ``return'' of this arm can be traced from \lb=$(0\Deg.9, 0\Deg)$ to
$(0\Deg.2, -0\Deg.07)$, and is more clearly recognized in Fig. 4
at $V=83 \sim 167$\kms.
This can be also clearly seen in the \lv\ diagram in
Fig. 2 at $b\sim -0\Deg.1$.
In the negative $l$ side, the arm appears to be bifurcated at
$l\sim -0\Deg.65$, and linked to Arm II.
This can be  more clearly observed in Fig. 4 at $V=-83 \sim 0$ \kms.
The intensity in Fig. 6a has been integrated to give a total mass of molecular
gas involved in GCA  I (in the velocity range as shown in Fig. 5a) to be
$M = 1.72  \times 10^7 \Msun $.

Arm  II can be traced as a a single bright ridge from $l\sim 0\Deg.3$
to $-0\Deg.7$, and the thickness is about 6$'$ (15 pc), and makes
GCA II.
The mass of Arm II is estimated to be
$ 1.35  \times 10^7 \Msun$.
Thus, the total mass involved in GCA  I and II is estimated to be
$ 3.07 \times 10^7 \Msun$, and shares almost 67\% of the total gas mass
in the observed region, and 78\% of the disk component.

\sub{3.2.2. The 120-pc Molecular Ring}

As shown in Fig. 4 and 6, GCA I and II compose a global ring structure,
which is  tilted and slightly bent.
If we fit the GCA I and II by a ring, its angular extent in the major axis
is $1\Deg.6$  from $l=-0\Deg.7$ to $0\Deg.9$,
and so, the major axis length (diameter) is 240 pc, and the
radius 120 pc.
The minor axis length is estimated  to be $7'.9$ from the maximum separation
between Arm I and II at $l\sim -0\Deg.2$ (see Fig. 7).
Therefore,  the inclination of the I+II ring is $i=85\Deg$ from the
minor-to-major axis ratio.
The center of the ring, as fitted by the above figures, is at
\lb=$(0\Deg.1,0\Deg.0)$
We call this ring the 120-pc Molecular Ring.

{}From these we conclude that the spatial distribution of the
molecular gas associated with the principal ridges in the \lv\ diagrams
comprises  a circum-nuclear ring of radius
$R\simeq 120$ pc inclined by 5\deg\ from the line of sight.

\sub{3.2.3. Cross Section of the Arms}

Fig. 7 shows the intensity variation perpendicular to the galactic
plane across GCA I and II averaged from $l=0\Deg.24$ to $-0\Deg.33$, where
the arms are most clearly separated.
Since the effective resolution of the present data is
$(\theta^2+\Delta b^2)^{1/2}=2'.0$, where $\theta=1'.7$ is the beam
width and $\Delta b=1'.0$ is the grid interval,
the arms are sufficiently resolved.
The two peaks in the figure at  $b=1'.8$ (Arm I) and  $b=-6'.0$ (Arm II)
can be fitted by a Gaussian intensity distributions as
$(T_{\rm B,~ peak}, {\rm FWHM}) = (0.27{\rm ~K}, 5'.3)$
and  $(0.33{\rm ~K}, 5'.5)$, respectively.
Namely, the arms are as thin as  13.0 (GCA I) and 13.5 pc (GCA II).
If we subtract the contributions from these two arm components, the
residual intensity in the whole area in Fig. 7 is only 36\% of
the total intensity.
The intensity coming from the inter-arm region between the arms shares
only 12\% of the intensity from the two arms.
This  would be an upper limit, as the region displayed in
this figure is the weakest part  along the arms without any
significant molecular clumps and condensations.
Thus, we conclude that the molecular gas as observed in the CO line
emission in the region discussed in this paper
is almost totally confined within the two major arms.
Therefore, the central 100 pc radius region is almost empty, making
a hole of molecular gas, except the nuclear few pc region surrounding Sgr A.

\ce{-- Fig. 7 --}

\sub{3.3. Velocity Field}

Fig. 8a shows a velocity field as obtained by taking the first moment
of the $(\Vlsr, l, b)$ cube, and therefore, an intensity-weighted
velocity field.
A general rotation characteristics is clearly seen along the major
axis of the ring feature at $b\simeq -6'$.
Sgr C molecular spur is seen as a negative velocity spur extending
 toward negative $b$.
GCA  III is seen as the tilted high-velocity plume at
$(l,b) \sim (0\Deg.2, 0\Deg.1)$.

In addition to these individual velocity structures, a large-scale
velocity gradient in the latitude direction is prominent in the sense
that the positive $b$ side has positive velocity and negative $b$ side
negative velocity.
This can be attributed to the fact that the high-velocity expanding
shell (ring) is more clearly seen in positive velocity at $b>0\Deg$, while
the negative velocity part more clearly at $b<0\Deg$ (see section 4).
This can be explained by a tilted nature of the expanding
oblate molecular shell, as will be discussed in section 4 based on an
analysis of \bv\ diagrams.
In fact, if we construct a velocity field, excluding the expanding ring
features, we obtain a rather regular velocity field as shown in Fig. 8b.

\ce{-- Fig. 8 --} % Velocity field

\sub{3.4. Possible Models for the Galactic Center Arms and Ring}

We here try to reproduce the \lv\ diagram based on a simple spiral
arm model.
According to the galactic shock wave theory (Fujimoto 1966; Roberts 1969)
and  the bar-induced shock wave theory (Sorensen et al 1976;
 Huntley et al 1978; Roberts et al 1979; Noguchi 1988; Wada and Habe 1992),
flow vectors of gas in the densest part along the shocked arms are almost
parallel to the potential valley that is rigidly rotating at a
pattern speed slower than the galactic rotation.
In such shocked flows, the gas cannot be on a closed orbit, but  is
rapidly accreted toward the center along deformed spirals.

As the simplest approach to simulate an \lv\ diagram, we assume
that the flow vector of gas is aligned along a spiral with a constant velocity
equal to the rotation velocity in the potential.
%The negative and positive velocities of Arms I and II at $l=0\Deg$,
%respectively, can be explained if the gas is spiraling into
%the center along spiral orbits.
%In this case, Arm I is far side and Arm II near-side of the nucleus.
%This configuration may be supported by the fact that a
%strong absorption of the  H$_2$CO line  is observed
%against the radio continuum of
%Sgr A (nucleus) at \vlsr=41.0 \kms\ with a large dispersion of
%$\Delta V=32.0$ \kms\ (Downes et al 1980), exactly coinciding with the
%velocity and width of Arm II.
Fig. 9a shows a  model, where we have assumed two symmetrical
spiral arms with a pitch angle $p=10\Deg$.
In addition to a constant circular rotation of gas ($\Vrot=$constant;
flat rotation curve), radial infall motion of $\Vrot {\rm sin}~p$ is
superposed, so that the gas is flowing along the arms into the central region.
The density distribution in the arms are shown by the spiral-like contours.
The azimuthally averaged density of gas has a hole at
the center, or it corresponds to a ring distribution of gas on which
two arms are superposed.
A calculated \lv\ diagram is shown by the superposed
contours with a tilted X shape.
The characteristic features in the observed \lv\ diagrams can be now
qualitatively reproduced.
Fig. 9b and 10c show cases where the spiral arms are oval in shape whose
major axis is inclined by $\pm 30\Deg$ from the nodal line.
Such a case may be expected when the oval potential or a bar in the
center is deep enough to produce a non-circular motion.
Fig. 9d-f are the same, but the density distribution along the arms
has the maximum at the center and the pitch angle is taken larger: $p=20\Deg$.
Again, the case of a circular rotation appears to reproduce the observation,
while the oval orbit cases result in more complicated \lv\ plots than the
observation.
Among these  models, the case shown in Fig. 9a or 9b appears to reproduce
the observed characteristics in the \lv\ plot (e.g. Fig. 3) reasonably well.
The model in Fig. 9d or 9e with the averaged gas density increasing toward the
center may explain observed Arms III and IV.
However, the cases corresponding to Fig. 9c and 9f may be excluded.

\ce{-- Fig. 9 --} % L-V simulation

According to the galactic shock wave theory  in a spiral
density wave or a bar potential,
the shocked gas  looses its azimuthal velocity so that
the \lv\ behavior becomes closer to the potential's pattern speed.
As the consequence, the apparent rotation velocity of the gas along
shocked spiral arms is  smaller than that from the rotation velocity.
This may be the reason why the observed maximum velocities of the
rings/arms (e.g., Arm I near Sgr B) are less than that expected from the
gravitational potential.

\sub{3.5. Deconvolution into Projection on the Galactic Plane:
 A Face-on View}

We may thus assume that the molecular gas is
on a ring or spiral arms whose pitch angle is not so large.
Then, it is possible to deconvolve the \lv\ diagram into a
spatial distribution in the galactic plane by assuming an
approximately circular rotation.
Thereby, we make use of the velocity-to-space transformation (VST),
which has been extensively applied to derive the HI gas distribution
in our Galaxy (Oort et al 1957).
Suppose that a gas element is located at a projected
distance $x~ (\simeq l\times 8.5$ kpc) along the galactic plane
from the center of rotation, and  has a radial velocity $v$.
If the rotation is circular at velocity $V_0$, the line of sight distance $y$
of the element from the nodal line can be calculated by
$$ y = \pm  |x| \sqrt{ \left( V_0 \over v \right)^2 -1} . \eqno(5)$$
The signs  must be opposite for Arms I and II.
Here, we assume that Arm I is near side, and Arm II far side, so that the
signs are  $-/+$, respectively.
The center of rotation is assumed to be at Sgr A, and
$v$ is measured from the intersection velocity at $l=-0\Deg.06$ on
each arm ridge.

Fig. 10 shows a  thus obtained ``face-on'' map of the molecular gas
for  $V_0=150$ \kms.
The arms appear to construct a
circum-nuclear ring of radius $\sim 120$  pc.
Here, we used \lv\ diagrams averaged within latitude ranges $-2' \le b \le 6'$
for Arm  I and $-5' \le b \le 3'$ for Arm II,
so that vertically extended clumps such as Sgr B complex are only
partly mapped in this figure.
During the deconvolution, we used only the arm component
concentrated near the ridges within $\pm 20$ \kms\ in velocity
(as illustrated in Fig. 5).
Diffuse gas and clumps with velocities far from the arms are not taken
into account.
The same VST was applied to the HII regions Sgr B1, B2 and C using their
H recombination line velocities (Downes et al 1980).
We plotted their positions  in Fig. 10.
The HII regions lie  along the arms associated with the molecular complexes,
though slightly avoiding the molecular gas peaks.
Sgr B and C appear to be at symmetrically opposite locations
with respect to the nucleus.
We have assumed that Arm I is near side. However, in this kind of
simple deconvolution, we cannot distinguish the exact orientation,
as is the case of deconvolution of gas distribution inside the
solar circle from kinematical information.
Hence, it may be possible to assume an opposite configuration of the arm
locations: Arm I in far side, and Arm II in near side.

\ce{-- Fig. 10 --}

The connection of Arms I and II is not clear from this deconvolution.
This is mainly because of the ambiguous position determination near the node,
which arises from unknown precise rotation curve.
The error is also large at $|l| < \sim 0\Deg.1$, where we
applied interpolation from both sides along  each arm.
Obviously, this kind of deconvolution is not unique, but it was possible here
because of the separation of Arms I and II in the $(l,b)$ plane.
Therefore, this deconvolution should be taken as a possible hint to
the spatial distribution of gas.

\sub{3.6. Comparison with Other  Galaxies and Models}

Accretion spirals, either shocked or not, and rings
of molecular gas have been indeed observed in the CO line
in many extragalactic systems such as IC 342 (Lo et al 1984;
Ishizuki et al 1990a) and  NGC 6946 (Ishizuki et al 1990b).
The ring  structure of molecular gas of 100 to a few
hundred pc size is  commonly observed in  the central
regions of spiral galaxies (Nakai et al 1987; Ishiguro et al 1989).
See  Sofue (1991) for a more number of galaxies with a nuclear molecular ring.
Thus, the ring/spiral structure of molecular gas of radius 120 pc in
the  Milky Way, would be similar to the situation found in external galaxies.

There have been various numerical simulations of the accretion of gas
toward the central region in spiral and oval potential
by  gas-dynamical simulations
(Sorensen et al 1976; Huntley et al 1978; Roberts et al 1979; Noguchi 1988;
Wada and Habe 1992).
The models predict a rapid accretion of gas along spiral orbits,
and the gas behavior in these models somehow mimic
the models illustrated in Fig. 9.

A number of simulations of position-velocity diagrams along the
galactic plane have been constructed and compared with the observations,
in order to understand larger-scale \lv\ diagrams for our Galaxy both in
HI and CO (Mulder and Liem 1986; Liszt and Burton 1978; Burton 1988).
Position-velocity diagrams for extragalactic edge-on galaxies
in CO have been extensively studied (Sofue and Nakai 1993, 1994; Sofue 1994)
and a numerical simulation has been attempted to reproduce the PV
characteristics based on the gas dynamics in an oval potential (e.g., Mulder
and Liem 1986; Wada et al 1994).

Binney et al (1991) have noticed the  ``parallelogram'' and
calculated theoretical \lv\ diagrams, and have shown
the presence of a bar of 2 kpc length in the Galactic bulge.
However, the parallelogram (the expanding ring feature) in Fig. 1b shares
only 15\% of the total emission.
However, we emphasize that the major structures,
which contain 85\% of the molecular mass within 150 pc of the center,
are due to the Arms discussed above.

\sub{3.7. Relationship with Radio Sources}

Fig. 11 shows superposition of a 10-GHz radio continuum map (Handa
et al 1987) on the $^{13}$CO and CS intensity maps.
We here briefly comment on a global relationship of the major
radio sources with  molecular features at a spatial resolution
of a few arc minutes.
Detailed  internal structures of individual sources are
out of the scope of the present paper, for which the readers may refer to
a review by Liszt (1988).

\sub{3.7.1. Sgr A}

The relationship of molecular features of scales less than
a few arc minutes with Sgr A has been discussed by many authors
(e.g., Oort 1977; Bally et al 1987; G{\"u}sten 1989).
However, these  nuclear features, which are of $1'~(\sim 3$ pc) scales,
are not well visible in the present plots in so far as the \lv\ plots
are concerned.
We only mention that Arm III is a largely tilted out-of-plane plume with
high  positive velocity, which Bally et al (1988) called the  polar arc.
Arm IV shows also large velocity gradient, and appears to be an object
related to a deep gravitational potential around the nucleus.

\sub{3.7.2.  Sgr B}

The molecular complex at $l\sim 0\Deg.6-0\Deg.9$ on  Arm I
is associated with the star forming regions Sgr B1
 at $(l,b)=(0\Deg.519,-0\Deg.050)$,  and Sgr B2
at $(0\Deg.670, -0\Deg.036)$.
Sgr B1 and B2, whose radial velocities in H recombination line emission are
\vlsr=45 and 65 \kms, respectively (Downes et al 1980),
are also located in the \lv\ plane at the upper (higher-velocity)
edges of molecular clumps.
Thus, the de-convolved positions of these continuum sources are slightly
displaced from the de-convolved arm, as indicated in Fig. 10.
The molecular gas distribution is highly extended in the direction
of latitude for about $0\Deg.4$ (60 pc), largely shifted toward the lower
side of the galactic plane ($b <0\Deg$).
This complex is also much extended in the velocity space:
the velocity dispersion amounts to as high as 50 \kms.
The internal structure of Sgr B molecular complex has been discussed
in detail in relation to the star formation activity, and it
was shown that the molecular gas is distributed in a shell, spatially
surrounding the continuum peak  (Bally et al 1988; Sofue 1990;
Hasegawa et al 1993).
The present ring model is consistent with the
CII line \lv\ diagram as obtained by Okuda et al (1989),
which indicates a rotating ionized gas feature with
Sgr B and C  on the tangential points of the ring.

\ce{-- Fig. 11 --}

\sub{3.7.3. Sgr C}

The star forming region Sgr C is associated with a molecular complex,
and is located on Arm II at $l\sim -0\Deg.6$.
However, the spatial proximity is less significant than that for Sgr B:
The radio continuum peak of Sgr C, $(l,b,\Vlsr)=(-0\Deg.57,-0\Deg.09)$,
 is located at the western edge of the molecular complex,
but  displaced by about 6$'$(15 pc) from the molecular peak.
The LSR velocity of the H recombination line also agrees with the
molecular gas velocity, and so, it is located on the de-convolved
arm in Fig. 10.
The molecular gas in this complex is  extended vertically, and
molecular spurs are found to extend both
toward positive and negative latitude directions.
We emphasize that the positive-latitude  spur is clearly associated with
the inner edge of the western ridge of the Galactic Center Lobe observed in
the radio continuum emission (Sofue and Handa 1984; Sofue 1985), as is
shown in Fig. 11.

\sub{3.7.4. Orbital Displacement vs Alignment of Star Forming Regions
and Molecular Arms}

The close association of Sgr B and C with GCA I and II
may have  a crucial implication for the orbits of gas and stars:
If the arms are shock lanes in a  bar during a highly non-circular
motion, the HII regions of a million  years old should already be
displaced from the molecular arms.
Therefore, the fact that Sgr B and C are still near the gas complexes
from which they may have been born (after one or more rotations) can be
explained only if the stars and gas are  circularly co-rotating in the
arms at a small pitch angle.
This would argue for the validity of the deconvolution process
applied in section 3.5.

Consider a spiral arm which is a shocked density wave.
Star formation from a molecular cloud will be triggered in the arms.
It will take about $t \sim 10^6$ years
for proto stars to form and shine as OB stars, and therefore,
until HII regions are produced.
On the other hand, the rotation period of the stars is
only $\sim 10^6$ years for $r=100$ pc and $V_{\rm rot}=200 $ \kms.
According to the density wave theory,  the velocity difference between the
rotation velocity and the shocked gaseous arm, which is about the same as
the pattern speed of density wave, is  of the order of
$$V_{\rm rot}-V_{\rm p}=(\Omega_{\rm rot}-\Omega_{\rm p})r. \eqno(6)$$
The azimuthal phase difference between  the HII
region and the gaseous  arm is then
$$\Delta \phi \sim (\Omega_{\rm rot}-\Omega_{\rm p})t . \eqno(7) $$

The phase difference for Sgr B2 and its corresponding molecular peak
in Fig. 10 (darkest part in Arm I) is roughly $\Delta \phi \sim 5\Deg$,
and a similar value is found for Sgr C.
If  $t\sim 10^6$ yr, we obtain
$\Omega_{\rm rot} - \Omega_{\rm p}\sim 0.1$ radian/$10^6$ years
$\sim 100$ \kms~kpc$^{-1}$.
This is an order of magnitude greater than the value near the solar circle
($\sim 10$ \kms~kpc$^{-1}$).
For older HII  regions (weaker radio sources) the phase difference
would be  much greater.
Moreover, orbits of stars, and therefore, HII regions, are no longer closed,
and must be largely displaced from the orbits of gas.
Thus, the HII regions in the central 100 pc
 of the Galaxy, except for young cases as Sgr B2,
would not be associated with molecular gas arms.
This will simply explain why the molecular gas features are not directly
correlated with the weaker radio sources in the Galactic center (Fig. 11).

\section{4. Discussion}

By analyzing the $^{13}$CO line BTL data cube, we have shown that
most (85\%) of the total molecular gas within $|l|<1\Deg$
comprises rigid-body-like structures
in the \lv\ diagrams, which can be  attributed to arms on a ring.
Moreover, 66\% of the total gas in the region, and 78\% of the disk component
($|b|<\sim 10'$=25 pc), was found to be confined in the two major
Arms I and II.
The spiral/ring structures are consistent with the picture drawn by Scoville
et al (1974) based on the earlier data, while the scale obtained here  is
slightly smaller.
The structures will be  common in external galaxy nuclei
in the sense that the gas distribution is  spiral- and ring-like.

Numerical simulations for a few kpc scale disks have suggested
that the features would be understood as the consequence of spiral
accretion by a density wave in an oval potential, either shocked or not.
Based on qualitative consideration,
we have suggested possible models to explain the observed \lv\ features
as shown in Fig. 9a.

The molecular mass in the Galactic Center has been derived
by usin the most recent CO-to-\htwo conversion factor
about one third of the conventional value, which
has been obtained by detailed analyses
of the dependency on the metallicity as well as on the galacto-centric
distance (Arimoto et al 1994).
This has resulted in a factor of three smaller mass and
energetics than the so far quoted values  in the literature:
The molecular mass within 150 pc radius from the center is estimated to be
only $3.9\times 10^7 \Msun$.

Thus, the molecular gas mass is only a few percent
of the total mass in the region estimated as
$M_{\rm dyn}=R V_{\rm rot}^2/G \sim 8 \times 10^8 \Msun$ for a radius
$R \sim 150$ pc and rotation velocity $V_{\rm rot} \sim 150$ \kms.
This implies that the self-gravity of gas is not essential in the
galactic center, and a given-potential simulation would be sufficient
to theoretically understand the region.

The expanding molecular ring (or the parallelogram) was shown to share
only 15 percent of the total gas mass within the central 1\deg\ region.
This feature has been shown to be extending vertically over $\sim 100$ pc
above and below the galactic plane (Sofue 1989).
For the very different $b$ distribution, it is a clearly
distinguished structure from the arms and the ring described in this paper.
On the \lv\ plot, the feature can be fitted by an ellipse of radius 1\deg.2
(Bally et al 1987), slightly larger than the disk discussed in this paper .
There have been controversial interpretations about this feature: either it
is due to some explosive event (Scoville et al 1972; Kaifu et al 1972, 1974)
or due to non-circular rotation of disk gas
(Burton and Liszt 1992; Binney 1991).
We will discuss this feature  based on the present data in a separate paper.

\vv
{\bf Acknowledgement}: The author would like to express his sincere thanks
to Dr. John Bally for making him available with the molecular line data
in a machine-readable format.

\section{References}

\parskip=0pt
\def\r{\hangindent=1pc \hangafter=1 \no}

\r Altenhoff, W. J., Downes, D., Pauls., T., Schraml, J. 1979, AAS {35}, 23.

\r Arimoto, N., Sofue, Y., Tsujimoto, T. 1994, in preparation.

\r{Bally, J., Stark, A.A., Wilson, R.W., and Henkel, C. 1987, ApJ Suppl 65,
13.}

\r{Bally, J., Stark, A.A., Wilson, R.W., and Henkel, C. 1988, ApJ 324, 223.}

%\r{Beichman, C.A., Neugebauer, G., Habing, H.J., Clegg, P.E., and Chester,
%%T.J. (ed.) 1985, IRAS Explanatory Supplement, JPL D-1855 (Jet Propulsion
%%Laboratory, Pasadena).}

\r Binney, J.J., Gerhard, O.E., Stark, A.A., Bally, J., Uchida, K.I., 1991
MNRAS 252, 210.

\r{Brown, R.L, and Liszt, H.S. 1984, ARAA 22, 223.}

\r Burton, W. B. 1988, in Galactic and Extragalactic Radio Astronomy,
ed. G. L. Verschuur and K. I. Kellermann, 2nd edition (Springer-Verlag,
New York) p 295.

\r Burton, W. B., and Liszt, H. S. 1983, in Surveys of the Southern Galaxy,
ed. W. B. Burton and F. P. Israel (Reidel Pub. CO, Dordrecht), p. 149.

\r Burton, W. B., and Liszt, H. S. 1992 AAS 95, 9.

\r Combes F 1992 ARAA, 29, 195.

\r Cox, P., Laureijs, R. 1989,
in The Center of the Galaxy (IAU Symp. 136),
ed. M.Morris (D.Reidel Publ. Co., Dordrecht) p. 121.

\r{Dame, T. M., Ungerechts, H., Cohen, R. S., de Geus, E. J., Grenier, I. A.,
et al. 1987 ApJ 32,  706

\r Downes, D., Wilson, T. L., Beiging, J., Wink,J. 1980, AA Suppl. 40, 379.

%\r Fabbiano, G. and Trinchieri, G. 1984, ApJ, 286,  491.

\r Fujimoto, M. 1966, in Non-stable Phenomena in Galaxies, IAU Symp. No 29,
ed. Arakeljan (Academy of Sciences of Armenia, USSR), p.453.

%\r{F{\"u}rst, E., Sofue, Y, and Reich, W. 1987, AA 191, 303.}

\r{Genzel, R., and Townes, C.H 1987, ARAA 25, 377.}

\r G{\"u}sten, R.  1989, in The Center of the Galaxy (IAU Symp. 136),
ed. M.Morris (D.Reidel Publ. Co., Dordrecht) p. 89.

\r{Handa, T., Sofue, Y., Nakai, N. Inoue, M., and Hirabayashi, H. 1987,
PASJ 39, 709.}

\r Hasegawa, T., Sato, F., Whiteoak, J. B., Miyawaki, R. 1993, ApJ 419, L77.

\r Heiligman, G. M. 1987 ApJ 314, 747.

\r{Huntley, J. M., Sanders, R. H., and Roberts, W. W.,  1978, ApJ 221, 521.}

\r{Ishiguro, M., Kawabe, R., Morita, K.-I., Okumura, S. K., Chikada, Y. et al.
1989, ApJ 344, 763. } %Maf 2.

\r{Ishizuki, S.  Kawabe, R., Ishiguro, M., Okumura, S. K., Morita, K-I.,
et al. 1990a Nature 344, 224. }%IC342

\r{Ishizuki, S., Kawabe, R., Ishiguro, M., Okumura, S. K., Morita, K. -I.
et al. 1990b ApJ 355 436. }%N6946

\r{Kaifu, N., Iguchi, T., and Kato, T. 1974, PASJ 26, 117.}

\r{Kaifu, N., Kato, T., and Iguchi, T. 1972, Nature 238, 105.}

\r Knapp, G. R., Stgark, A. A., Wilson, R. W. 1985 AJ 90, 254.

%\r{Koyama, K., Awaki, H., Kunieda, H., Takano,S., Tawara, S., Yamanuchi, S.,
%Hatsukade, I., and Nagase, F. 1989, Nature  339, 603.}

\r Liszt, H. S. 1988 in Galactic and Extragalactic Radio Astronomy,
ed. G. L. Verschuur and K. I. Kellermann, 2nd edition (Springer-Verlag,
New York) p 359.

\r Liszt,H. S., Burton, W. B. 1978 ApJ 226, 790.

%\r Liszt,H. S., Burton, W. B. 1978 ApJ 236, 779.

\r Liszt, H. S., and Burton, W. B. 1980 ApJ 236, 779.

\r{Lo, K. Y., Berge, G. L., Claussen, M. J., et al. 1984, ApJ 282, L59.}%IC342

\r Mulder, W.A., Liem, B.T., 1986, AA 157, 148

\r{Nakai, N., Hayashi, M., Handa, T., Sofue, Y., Hasegawa, T., and Sasaki, M.,
1987, PASJ  39, 685.}

\r Okuda, H., Shibai, H., Nakagawa, T., Matsuhara, T., Maihara, T., et al.
1989 in The Center of the Galaxy (IAU Symp. 136),
ed. M.Morris (D.Reidel Publ. Co., Dordrecht) p. 145.

\r Oort, J. H., Kerr, F. J., Westerhout, G. 1958, MNRAS 118, 379.

\r Oort, J. H. 1977, ARAA  15, 295

%\r{Reich, W., F{\"u}rst, E., Steffen, P., Reif, K., and Haslam, C.G.T. 1984,
AA Suppl 58, 197.}

%\r{Reich, W., Sofue, Y., and F{\"u}rst, E. 1987, PASJ 39, 573.}

\r Roberts, W. W. 1969, ApJ 158, 123.

%\r Roberts, W. W., Yuan, C. 1970, ApJ 161, 877.

\r Roberts, W. W., Huntley, J. M., van Albada, G. D. 1979, ApJ, 233, 67.

%\r{Saito, M., and Saito, Y. 1977, PASJ 29, 387.}

\r Shaver, P. A., McGee, R. X., Newton, L. M., Danks, A. C., Pottasch, S. R.
1983, MNRAS, 204, 53.

\r{Scoville, N.Z. 1972, ApJ  175, L127.}

\r Scoville, N.Z., Solomon, P. M., and Jefferts, K. B. 1974 ApJ 187, L63.

%\r{Sofue, Y. 1984, PASJ 36, 539.}

\r Sofue, Y. 1985 PASJ 37, 697

\r Sofue, Y. 1989, in The Center of the Galaxy (IAU Symp. 136),
ed. M.Morris (D.Reidel Publ. Co., Dordrecht) p. 213.

%\r Sofue, Y.  1989b Ap. Let.  Com. 28, 1 %cylindrical 200-pc ring

\r Sofue, Y. 1990 PASJ  42, 827 % Radio Continuum and CO line Emissions in

\r Sofue, Y. 1991 PASJ  43, 671 %Mol. ring and distance scale

\r{Sofue, Y., and Handa, T. 1984, Nature 310, 568.}

\r Sofue, Y., Nakai, N. 1993 PASJ 45, 139. %N891

\r Sofue, Y., Nakai, N. 1994 PASJ 46, 147.%N4565

\r{Sofue, Y., and Reich, W. 1979, AA Suppl  38, 251.

\r{Solomon, P.M., Scoville, N.Z., and Sanders, D.B., 1979, ApJ  232, L89.}

\r{S$\phi$rensen, S. -A., Matsuda, T., and Fujimoto, M. 1976,  A. Sp. Sci.
43, 491. }

\r Stark, A. A., Bally, J., Wilson, R. W., Pound, M. W., 1989,
in The Center of the Galaxy (IAU Symp. 136),
ed. M.Morris (D.Reidel Publ. Co., Dordrecht) p. 213.

\r Tsuboi, M. 1989 in  The Galactic Center (IAU Symp. 136),
ed. M.Morris (D.Reidel Publ. Co., Dordrecht) p. 135

\r Wada, K., Habe, A., Taniguchi, Y., Hasegawa, T. 1994, submitted to Nature

\r Wada, K., Habe, A. 1992 MNRAS 258, 82

\endpage
\hsize=180truemm
\vsize=250truemm
\settabs 7 \columns

\no Table 1: Galactic Center Arms and Ring.
\v
\hrule
\vskip 0.4mm \hrule
\v
\+ Parameters & & Ring (I+II)& Arm I & Arm II&  Arm III& Arm IV   \cr
\v
\hrule
\v
\+ From \lv&(\deg, \kms) \dotfill &(+0.9,90) & $(0.9,80)$ & $(0.1,60)$
&$(0,140)$ &$(0.1,60)$ \cr
\+& & & & $\sim(1,100)$ & & \cr
\+ To \lv&(\deg, \kms) \dotfill & $(-0.65,-140)$ &$(-0.7, -150)$ &$(-0.6,-80)$
&$(-0.15,10)$ & $(0,-20)$ \cr
\+& & & $\sim(-1,-200)$ & & & \cr
\+ \vlsr at $l=0\Deg$&(\kms) \dotfill &~~~~.... &$-40$  &+50 &+70 & $-50$ \cr
\v
\+ From \lb&(\deg,\deg)\dotfill &$(+0.9,0.0)$ &$(+0.9,-0.1)$ & $(0.25,-0.05)$
& $(0.25,0.25)$ &~~~~.... \cr
\+ To \lb&(\deg,\deg) \dotfill & $(-0.65,-0.08)$& ~$(-1.0,-0.2)$
&$(-0.65,-0.17)$ &$~(0,0)$ & ~~~~....\cr
\+ $b$ at $l=0\Deg$&(\deg) \dotfill &~~~~.... & 0.050& $-0.067$&$(0,0)$
&~~~~.... \cr
\+ Length &(\deg/pc) \dotfill &~~~~.... & 1.9/280 & 0.9/133 &
0.35/52&~~~~....\cr
\+ Min. $b$ width&(\deg/pc)\dotfill &~~~~.... &0.088/13 & 0.091/13.5&~~~~....
&~~~~....\cr
\+ Max. $b$ width&(\deg/pc)\dotfill &~~~~....& 0.33/50 &
0.2/30&~~~~....&~~~~.... \cr
\vv
\+ Maj.ax.len.&(\deg/pc)\dotfill &1.55/230 & ~~~~.... &~~~~.... &
{}~~~~....&~~~~.... \cr
\+ Min.ax.len.&(\deg/pc)\dotfill &0.132/19.5 & ~~~~.... &~~~~.... &~~~~....
&~~~~.... \cr
\+ Inclination&(\deg)\dotfill &$85\Deg.1$ & ~~~~.... &~~~~.... &~~~~....
&~~~~.... \cr
\+ Ring cen. \lb& ~~(\deg,\deg) \dotfill &(0.12,0.0) & ~~~~.... &~~~~....
&~~~~....
&~~~~.... \cr
\+ Ring radius &  (pc) \dotfill & 120  &  & & & \cr
\+ Rot. Velo &(\kms)\dotfill &$+90/-140$ & ~~~~.... &~~~~.... &~~~~....
&~~~~.... \cr
\vv
\+ Mol. Mass$^\dagger$& ($10^7\Msun$)\dotfill & 3.07  & 1.72  & 1.35  &
{}~~~~....& ~~~~.... \cr
\+ Remarks & \dotfill & Circum Nuc.& asso. Sgr B & Sgr C & Sgr A? & Sgr A?\cr
%\+ & \dotfill & &  & & & \cr

\v
\hrule
\v

\no $*$ The distance to the galactic center is assumed to be 8.5 kpc.

\no $\dagger$ 1.61 times the \htwo\ mass obtained from the
$^{13}$CO intensity to \htwo\ conversion [see eq. (1)-(3)],
where the metal abundance has been assumed to be twice the solar.
This also applies to mass in Table 2.
The statistical error which occurs during intensity integration
is only a few \%,
while the error arising from  ambiguity of the conversion factor is
about 20 to 30\%  (Arimoto et al 1994).

\endpage
\section{Figure Captions}

Fig. 1: (a) The \lv\ diagram of the $^{13}$CO $(J=1-0)$ line emission
of the central region of the Milky Way by averaging the data from
$-0.35 \le b \le 0\Deg.17$
as obtained with the Bell Telephone 7-m telescope by Bally et al (1987).
Contours are in unit of K \ta\ at levels
$0.1\times$(1, 2, 3, 4, 5, 6, 8, 12, 15, 20, 25).

(b) The same as Fig. 1a,  but the local and foreground CO emissions
have been subtracted by applying the ``pressing method'' (see the text
for the procedure). Contour levels are same as in (a).

\vv
Fig. 2: The \lv\ diagrams averaged in $4'~b$ interval.
Local/foreground emissions have been removed.
`Rigid-rotation' ridges (arms) are dominant in the disk
at $|b|<\sim 10'$ (25 pc).
Contours are in unit of K \ta\ at levels
$0.2\times$(1, 2, 3,..., 9, 10, 12, 14, 16, 18, 20, 25, 30).

\vv
Fig. 3: Schematic sketch of the major ridges (arms) in the \lv\ diagrams.

\vv
Fig. 4: (a) Integrated intensity map in the whole velocity range at
$-250 \le \Vlsr \le 250$ \kms. This is almost the same as the map
presented by Stark et al (1989), except that the local contribution has
been subtracted.
Contours are in unit of K \kms\ at levels
$25 \times$(1, 2, 3, ..., 9, 10, 12, 14, 16, 18, 20, 25, 30).

(b) Same but in a grey-scale representation. For intensity scale, see (a).
The bottom figure shows the same, but the scale in the latitude direction
has been doubled.
Galactic Center Arm (GCA) I runs as a long arc in the positive $b$ side;
GCA II runs in the negative $b$ side.

\vv
Fig. 5: \lv\ diagrams  corresponding to
(a) Galactic Center Arms I and (b) II, which were used to obtain
intensity maps of the Galactic Center Arms in Fig. 6.
Contours are in unit of K \ta\ at levels
$0.2\times$(1, 2, 3, 4, 5, 6, 8, 10, 12, 15, 20, 25, 30, 35).

\vv
Fig. 6: Integrated intensity maps corresponding to (a) Galactic Center Arm I,
and (b) Arm II as in Fig. 5. Contours are in unit of K \kms\ at levels
$12.5\times$(1, 2, 3, ..., 9, 10, 12, 14, 16, 18, 20, 25, 30).
%(cu 0.025x500 km/s).
(c) Arms I+II. Contours are in unit of K \kms\ at levels  $25\times$(as above).

\vv
Fig. 7: Intensity variation across  Galactic Center Arms
I and II perpendicular to the galactic plane averaged at
$l=0\Deg.24$ to $-0\Deg.33$, where the arms are most clearly separated.

\vv
Fig. 8: (a) A velocity field as obtained by taking the first moment of
the $(\Vlsr, l, b)$ cube (intensity-weighted mean velocity field).
Contour interval is 10 \kms\. Full-line contours are for positive velocity
starting at 0 \kms. Dashed contours are for negative velocity.

(b) Same as (a), but for the ``disk component'' with $|\Vlsr|<100$ \kms.

\vv

Fig. 9: Two-armed spiral model with a spiral infalling motion.
Gas density distribution is shown by spiral-like contours as projected on
the galactic plane.
Calculated \lv\ diagram is shown by tilted X shaped contours.
The scales are arbitrary.

(a) Two  spiral arms with a pitch angle $p=10\Deg$ are assumed.
The azimuthally averaged gas density has a hole at
the center, corresponding to a ring distribution of gas on which
two arms are superposed.
In addition to a constant circular rotation,
radial infall of velocity $\Vrot {\rm sin}~p$ is superposed.

(b), (c) The same as (a), but  the spiral arms are oval in shape whose
major axis are inclined by $\pm 30\Deg$ from the nodal line.

(d)-(f) The same as (a)-(c), respectively,
but the density distribution along the arms
has the maximum at the center and the pitch angle is taken larger: $p=20\Deg$.

\vv
Fig. 10: Possible deconvolution of the \lv\ diagrams for Galactic Center
Arms I and II into a spatial distribution  as projected on the galactic plane.
Contour interval is 0.25 starting at 0.1 in an  arbitrary unit.
Sgr A is assumed to be at the center.

\vv Fig. 11: Superposition of the radio continuum emission at 10 GHz (contours:
Handa et al 1987) on (a) $^{13}$CO, and (b) CS emission maps (grey scale).
Contours are in unit of K \tb\ of 10 GHz continuum brightness at levels
$0.1\times$(1, 2, 3, 4, 6, 8, 10,  15, 20, 25).
For CO intensity scale, see Fig. 4.

\bye